%Definitions

\def\({\left(}
\def\){\right)}
\def\[{\left[}
\def\]{\right]}

\def\a{\alpha}
\def\be{\beta}
\def\g{\gamma}
\def\G{\Gamma}
\def\de{\delta}

\def\la{\lambda}

\def\f{\phi}
\def\F{\Phi}
\def\om{\omega}
\def\Om{\Omega}

\def\vf{\varphi}

\def\S{\Sigma}

\def\ext{\land}
\def\coma{\quad ,\quad}

\def\lora{\longrightarrow}

\def\frac#1#2{{#1\over#2}}
\def\dfrac#1#2{{\displaystyle{#1\over#2}}}
\def\pd#1#2{\dfrac{\partial#1}{\partial#2}}
\def\dt#1#2{\dfrac{d#1}{d#2}}

\def\vspace#1{\crcr\noalign{\vskip#1\relax}}

\def\vs2{\vspace{2mm}}

\magnification= \magstep1
\hsize=4.9in
\vsize=8in
\hoffset=14truemm
\baselineskip 14 pt
\overfullrule 1pt
%\nopagenumbers

\centerline{\bf ON GYROSCOPIC PRECESSION}
\medskip
\centerline{J. L. Hern\'andez-Pastora$^\dagger$, J. Mart\'\i
n$^\dagger$, E. Ruiz \footnote{$^\dagger$}{Area de F\'\i sica Te\'orica.
 Edificio Triling\" ue, Universidad de Salamanca. 37008 Salamanca, Spain} }

\medskip

\vskip  2mm

{\bf Abstract}

The vorticity of a congruence is often considered to be  the rate of
 rotation for the precession of a gyroscope moving along a world-line belonging to that congruence. Our aim here was to determine the evolution equation for the angular momentum of a gyrosocope with respect to an arbitrary  time-like congruence: i.e, a reference congruence  not containing the curve described by the gyroscope. In particular, we  show the specific conditions
needed to support the introductory statement about the vorticity. We thus
establish a well-founded theoretical description for the analysis
of the precession of gyroscopes, providing suitable conclusions for
possible experiments.

\vskip 1cm

\noindent {\bf 1. INTRODUCTION}

\vskip 0.5cm

Analysis of the evolution undergone by the angular momentum
of a gyroscope in the presence of a gravitational field is a
problem in General Relativity that has reignited the interest aroused
 by the  expected (although delayed) launching
of the artificial satelite {\it Gravity Probe B} (GP-B) [1]. Thus, in recent 
 years a series of articles [2]-[8] studying
different aspects of this issue has been published, showing 
explicitly that despite the time elapsed since the early pioneering
works on the topic [9]-[12], there are still some obscure aspects that require further elucidation.

Among such articles, 
one by Rindler and Perlick [2] is outstanding. That paper establishes the foundations for studying the evolution of a point-like gyroscope
(i.e., of negligible size) with respect to a reference
system considered as a congruence of time-like world-lines. This
point of view is essential in order to gain both a model of the 
structure supporting the gyroscope and any  other reference
that  can be used to evaluate the physical magnituds involved in
the problem. In this paper we propose a specific procedure that allows us to recover the classical
post-newtonian results of  Fokker--de Sitter and Schiff in a fairly simple way.

In particular, we review  article [2] because it explicitly calculates  the orientation of the gyrosocope (changes in a certain angle) after its revolution along the orbit and -as explained below- this is the result that we are specifically interested in.   To be fair, however, other works should also be mentioned. In this sense, Massa and Zordan [13] conducted a rigorous study of the motion of a gyroscopically stabilized point compass in a given frame, applying the theory of space tensors. To a certain extent, those authors managed to modernise the Cattaneo-Zel'manov [14-17] approach to the splitting of space--times. Later on,  exhaustive work by Jantzen, Carini and Bini [18] updated these techniques within the framework of the spatial gravitational forces used to treat gyroscope precession. In two papers [6],[19], one of those authors (D. Bini) devoted extensive space to studying gyroscope evolution and spinning test particles in general relativity, including the effects of gravitational waves effects on the  behaviour of gyroscopes.

The aim of the present work is to offer a complementary analysis with 
respect to the above-mentioned study by Rindler and Perlick
[2] as regards two main aspects. First, we wished to analyse the evolution of a gyroscope's angular momentum with respect to an arbitrary
congruence; i.e.,  a congruence which does not contain
the curve along which the gyroscope is moving. Thus, our aim is to show how the intrinsic quantities defined by the congruence on the quotient manifold (acceleration or Newtonian field, rotation or Coriolis field and, deformation
rate) affect the evolution of the gyroscope,  as done in [13]. In particular, we obtain the key equation of  Rindler and Perlick  when the gyroscope is assumed to be at rest with respect to the congruence, which, additionally, has no deformation rate (Born's congruence). As stated by the authors, this result implies that the precession rate of the gyroscope coincides with the rotation tensor of the congruence.

 Second, we wished to determine the change in the  orientation of a gyrscope after one revolution in its orbit as appreciated by some arbitrary observer. To do so, we make use of  a triad of connectors transported by paralelism in the sense of the spatial metric and we take into account the holonomy group of such connectors transported along closed curves in a Riemannian space.

In [2], the authors are mainly interested in the final rotated angle made by the gyrosocope after a given period in a closed orbit and do not take into account the influence of gravitational fields associated with an arbitrary  congruence. Accordingly, the problem about the change from the ``proper" congruence to an arbitrary one can be solved intuitively by taking the standard coordinates of the Schwarzschild metric (and the
Boyer--Lindquist coordinates of the Kerr metric) at the level of the spherical
coordinates of  Minkowski flat space. We thus assume that  the tangent vectors to the radial coordinates rotate
an angle of $2 \pi$ when a  displacement of one revolution is made
along a circular orbit on the equatorial plane. We propose the prior definition of a new angular
momentum by means of a suitable boost.  This approach has  already been used by Bel and one of us, (Mart\'\i n) [20], to transform a spin vector from one frame of reference to another within the context of the predicitive relativistic mechanics of systems with spin, and it has also been used more recently by Bini [6] to  study the evolution of gyroscopes. We then evaluate the
rotation of this spin (new angular momentum) with respect to a
triad that is transported by parallelism in the sense of the
quotient metric associated with the congruence. Finally, we see that it is
possible to recover the results of Rindler and Perlick if
the proper rotation of the triad is considered for a complete
revolution of the circular orbit.

The work is  presented as follows. Section  2 offers a brief review of the time-like congruences of space-time, placing emphasis on certain aspects to be deployed later on, such as the concept of natural and orthogonal connector triads.

Section 3 is divided into three subsections. In the first, we explore the evolution of a gyroscope with respect to a congruence and the construction of a new spin by means of a local Lorentzian boost. In the second, we obtain the evolution equation of the new spin in the three-dimensional formalism associated with  the reference congruence. Finally, in the third subsection we introduce  a field of non-natural connector triads  in order to generalise the Cartesian connectors of Minkowski space-time.

Finally, in order to illustrate this issue, in
section   4  we evaluate the precession for Schwarzschild and Kerr metrics. Our aim in doing this is not only to show that this methodology  provides
the  correct (and  well known) results but also to highlight the 
difficulties that would be encountered on handling this 
problem in more general cases.

\vskip 5mm

\noindent {\bf 2. TIME-LIKE CONGRUENCES}

\vskip 2mm

In this section we explicitly   introduce the
definitions to be used later on regarding a   time-like
congruence. Let ${\cal C}$ be a congruence of time-like world-lines
on some domain ${\cal D}$ of the space-time manifold  (${\cal
V}_4\,, g_{\a \be}$), and let
$$
x^\a = f^\a(p\,, z^i)
\eqno(1)
$$
($i,j,...=1,2,3 ;\ \a,\be,...=0,1,2,3$) be the parametric equations
of this congruence. The unit time-like tangent vector field is:
$$
u^\a(x) = \xi^{-1}\,\xi^{\a}[p(x), z^i(x)] \coma g_{\a\be}u^\a
u^\be = -1 \,,
\eqno(2)
$$
with
$$
\xi^\a \equiv \pd{f^\a}{p}\coma g_{\a\be}\xi^\a\xi^\be \equiv -\xi^2 < 0 \,,
\eqno(3)
$$
where
$$
\left\{\eqalign{
p &= p(x^\a) \cr
\vs2
z^i &= z^i(x^\a) \cr
}\right.
\eqno(4)
$$
are the inverse  functions of (1). The time-like parameter $p$ of
the congruence can be choosen, up to the gauge transformation
$\tau_{_R} \to \tau_{_R} + A(z^i)$, to be the proper time
of the congruence,  such that:
$$
\tau_{_R}:\ u^\a=\pd{x^\a}{\tau_{_R}} \,.
\eqno(5)
$$

The three tensor fields associated intrinsically with the time-like
congruence - the Deformation rate $\S_{\a\be}\,$, the Rotation
$\Om_{\a\be}$ ( sometimes called the  Coriolis force or Gravitomagnetic
field) and the Acceleration $ b_\a$ (Newtonian or Gravitoelectric field
up to sign)- are given by:
$$
\nabla_\a u_\be = \S_{\a\be} + \Om_{\a\be} - u_\a b_\be \,,
\eqno(6)
$$

$$
\left\{\eqalign{
\S_{\a\be} &= \frac12 \hat g^\la_\a\,\hat g^\mu_\be(\nabla_\la
u_\mu + \nabla_\mu u_\la) \cr
\vs2
\Om_{\a\be} &= \frac12 \hat g^\la_\a\,\hat g^\mu_\be(\nabla_\la
u_\mu - \nabla_\mu u_\la) \cr
\vs2
b_\a &= u^\rho\nabla_\rho u_\a \,, \cr
}\right.
\eqno(7)
$$
where
$$
\hat g^\la_\a \equiv \de^\la_\a + u^\la u_\a
\eqno(8)
$$
is the projector tensor orthogonal to $u^\a$ (usually considered to be 
the metric on the 3-space quotient manifold).

Since we are interested in the evolution of a vector orthogonal
to $u^\a$, to refer to it  we first  define a 3--frame orthogonal to $u^\a$.
 The natural way to do this is as follows: since each
line of the congruence is characterized by a parameter $z^i$, a
vector connecting two lines of (1) and orthogonal to $u^\a$ is
given by a linear combination (with functions independent of the
parameter
$p$) of  the spatial projection of the three derivatives:
$$
Q^\a_i \equiv \pd{f^\a}{z^i} \,.
\eqno(9)
$$
i.e.
$$q^\a_i \equiv \hat
g^\a_\la\,Q^\la_i = \hat g^\a_\la\,\pd{f^\la}{z^i}\, \,,
\eqno(10)
$$
which we shall henceforth refer to as the {\it natural orthogonal
connectors triad} $\{q_i^\a\}$. These connectors are the components of the
following vector-field:
$$
\hat\partial_i \equiv q^\a_i\pd{\ }{x^\a} = \vf_i\pd{\ }{p} +
\pd{\ }{z^i} \coma \vf_i \equiv \xi^{-1}(u\,Q_i)\,.
\eqno(11)
$$

Thus, a tetrad of space-time is given by $\{e^\a_a\} \equiv \{u^\a,
q^\a_i\}$, ($a,b,...=0,1,2,3$), which satisfies the orthogonality
conditions:
$$
g_{\a\be}\,u^\a\,q^\be_i = 0 \,,
\eqno(12)
$$
where
$$
g_{\a\be}\,q^\a_i\,q^\be_j \equiv \hat g_{ij}
\eqno(13)
$$
are the components of the quotient metric with respect to the triad
$\{q^\a_i\}$. The related co-base of one-forms can be constructed
to give $\{\theta_\a^a\} \equiv \{-u_\a\,,\,p_\a^i\}$, such that:
$$
\left\{\eqalign{
&p_\a^idx^\a = \pd{z^i}{x^\a}dx^\a = dz^i \cr
\vs2
&q^\a_i\,p_\a^j = \delta^i_j \,.
\cr }\right. 
\eqno(14)
$$

It is worthwhile noting the behaviour of  orthogonal
connectors triads under a change in the parameters of the congruence
(1). If a general change is performed, such as:
$$
\left\{\eqalign{
p &\to p' = p'(p,z^i) \cr
\vs2
z^i &\to z^{k'} = z^{k'}(z^i) \,, \cr
}\right.\ 
\eqno(15)
$$
one then has:
$$
Q^\a_i = \xi' u^\a \pd{p'}{z^i} + Q^\a_{k'}\pd{z^{k'}}{z^i} \,,
\eqno(16)
$$
with
$$
\xi' = \(\pd{p'}{p}\)^{\!\!-1}\xi \,.
\eqno(17)
$$
Therefore, the natural orthogonal connector changes like a
tensor on the quotient manifold, whatever the time parameter $p'$
considered:
$$
q^\la_i = q^\la_{k'}\pd{z^{k'}}{z^i} \,.
\eqno(18)
$$

For further calculations it will be useful to obtain  the
coefficients $\g^c_{ab}$ of the connection with respect to the
tetrad $\{e^\a_a\} \equiv \{u^\a, q^\a_i\}$ in terms of the
geometrical objects defining the congruence. As is well known, we
have:
$$
e^\la_b\nabla_\la e^\mu_a = \g^c_{ab}\,e^\mu_c \,.
\eqno(19)
$$
By contrast, we also have:
$$
d \theta^c = -\frac 12 C^c_{ae} \theta^a \wedge \theta^e \,,
\eqno(20)
$$
$C^c_{ae}$ being  the coefficients appearing in the Lie brackets of the 
tetrad vectors; i.e.,
$$
[\vec e_a , \vec e_b ] = C^c_{ab} \vec e_c \,.
\eqno(21)
$$

Now using the definition of connectors, we obtain:
$$
\left\{\eqalign{
&\[\vec u\,,\vec q_j\] = (b\,q_j)\,\vec u \cr
\vs2
&\[\vec q_i\,,\vec q_j\] =2\,\hat\Om_{ij}\,\vec u \,, \cr
}\right.
\eqno(22)
$$
and it is therefore  easy to derive the following expressions for the
coefficients of the connection:

$$
\left\{\eqalign{
&\g^0_{00} = 0  \coma \g^0_{0k} = 0 \coma \g^0_{j0} = (b\,q_j)
\coma \g^0_{jk} = \hat\S_{kj} + \hat\Om_{kj} \cr
\vspace{4mm}
& \g^i_{00} = \hat b^i \coma \g^i_{j0} = \g^i_{0j} = \hat\S_j^{\ i}
+ \hat\Om_j^{\ i} \coma \g^i_{jk}
= \g^i_{kj} \equiv \hat\G^i_{jk} \,, \cr
}\right.
\eqno(23)
$$
where the following notation has been used (the latin indices are
raised and lowered with the quotient metric $\hat g_{ij}$):
$$
\left\{\eqalign{
\hat b_i &\equiv q^\a_i\,b_\a \cr
\vs2
\hat\Om_{ij} &\equiv
q^\a_i\,q^\be_j\,\Om_{\a\be} = \frac12\xi(\hat\partial_i\vf_j -
\hat\partial_j\vf_i) \cr
\vs2
\hat\S_{ij} &\equiv q^\a_i\,q^\be_j\,\S_{\a\be} =
\frac12\xi^{-1}\pd{\hat g_{ij}}{p} \,,
\cr
}\right.
\eqno(24)
$$
where $\hat\G^i_{jk}$ is the Zel'manov-Cattaneo connection
[16],[21]-[22]:
$$
\hat\G^i_{jk} = \frac12\hat g^{ih}(\hat\partial_j\hat g_{kh} +
\hat\partial_k\hat g_{jh} - \hat\partial_h\hat g_{jk}) \,.
\eqno(25)
$$

\vskip 1cm

\noindent {\bf 3. THE EVOLUTION OF A GYROSCOPE WITH RESPECT TO AN
ARBITRARY CONGRUENCE}

\vskip 0.5cm

Let us now consider a point-like gyroscope (of negligible size) moving along an arbitrary time-like curve. As usual, we can dispense with the Papapetrou equations [23] and assume that the spin of the gyroscope is Fermi-Walker-transported along the curve. The aim of this section is to provide an exact description of  the evolution of this spin with respect to a certain reference time-like congruence, and hence in terms of the geometric objects associated with  it: acceleration, rotation, deformation rate and Cattaneo's connection. This generalizes, for example, the conclusions
concerning the postnewtonian experiment of a rocket gyroscope in 
orbit around the Earth [24]. In addition, as we will see, the question of recovering  the well known Thomas or Fokker--de
Sitter and Schiff postnewtonian precession terms is not a trivial matter, because evaluation of the  angle rotated by the spin of the gyroscope is
not an issue  of superficial geometric considerations.

\vskip 2mm

\noindent {\bf A) Fermi-Walker transport and local covariant
Lorentzian boost}

\vskip 2mm

Let $S^{\a}$ be a Fermi-Walker-transported (FWT) vector along a
time-like curve with  a unitary tangent vector $w^{\a}$ and an acceleration of
$a^\a$:

$$
x^\a = \vf^\a(\tau) :\quad
\left\{\eqalign{
& w^\a \equiv \dt{\vf^\a}{\tau} \cr
\vs2
& a^\a \equiv
\frac{\nabla w^\a}{d\tau}  \cr
}\right. \coma g_{\a\be}w^\a w^\be = -1 \,,
\eqno(26a)$$

$$
\frac{\nabla S^\a}{d\tau} + (a^\a\,w_\la -
a_\la\,w^\a)\,S^\la = 0 \,.
\eqno(26b)
$$
The first two  integrals of this differential equation are given by
$S_{\mu} w^{\mu}$ and the length of $S^{\a}$, i.e.
$g_{\a\be}S^\a S^\be$. Therefore, we can take $S_{\mu} w^{\mu}=0$,
and so from here onwards $S^{\a}$  will represent the intrinsic spin
vector of a gyroscope moving along such a curve.

Let ${\cal C}$  now be a congruence of time-like world-lines defined
as usual:
$$
{\cal C}:\ x^\a = f^\a(\tau_{_R}, z^i)\coma \left\{\eqalign{
\tau_{_R} &= \tau_{_R}(x^\a) \cr
\vs2
z^i &= z^i(x^\a) \,,\cr
}\right.
\eqno(27)
$$
$\tau_{_R}$ being the proper time and $u^\a$ the unit tangent vector
$$
u^\a(x) = \pd{f^\a}{\tau_{_R}}[\tau_{_R}(x), z^i(x)] \coma
g_{\a\be}u^\a u^\be = -1 \,.
\eqno(28)
$$
An observer evolving with the reference congruence sees the
gyroscope in motion  because in general $u^\a$ is not co-linear with
$w^\a$. One therefore needs to know what  the components of the
spin vector $S^\a$ are as seen by this observer. To do so, we define
a covariant local Lorentzian boost  $ {\cal B}:\ \{u^\a\} \lora
\{w^\a\}$, thereby transforming this observer at rest for the congruence 
into an observer tied to the gyroscope. By applying the boost to
the spin-vector $S^\a$, we have:
$$
S^\a \,\lora\, N^\a = S^\a + \frac{(Su) + (Sw)}{1-(uw)}(u^\a +
w^\a) - 2(Sw)u^\a \,.
\eqno(29)
$$
This transformation  fulfills the following properties:
$$
\left\{\eqalign{
(Nu) &= (Sw) \cr
\vs2
N^2 &= S^2 \cr
\vs2
(Nw) &= - (Su) - 2(Sw)(uw) \,.\cr
}\right.
\eqno(30)
$$
In particular, and as  shown, this preserves the length of the
spin-vector. Moreover, because  the spin-vector $S^\a$ is
orthogonal to $w^\a$, we have:
$$
\left\{\eqalign{
(Nu) &= 0 \cr
\vs2
(Nw) &= - (Su) \,, \cr
}\right.
\eqno(31)
$$
and hence the new spin vector $N^\a$ is orthogonal to the reference congruence.

Let us now write the evolution equation of the transformed
spin-vector $N^\a$ along the curve $\vf^\a(\tau)$. Since $S^\a$
satisfies the FWT equation (26b), from (29)  it is straightforward to obtain:
$$
\eqalign{
\frac{\nabla N^\a}{d\tau} = &\frac{(Na)}{1+\g}(w^\a - \g u^\a) -
\frac{(Nw)}{1+\g}\[a^\a + (ua)u^\a\] \cr
\vs2
&+ \frac1{1+\g}\[N_\rho(u^\a+w^\a) - (Nw)\delta_\rho^\a\]
\frac{\nabla u^\rho}{d\tau} \,, \cr
}
\eqno(32)
$$
with
$$
\frac{\nabla u^\rho}{d\tau} = w^\la(\S_\la^{\ \a}+\Om_\la^{\ \a} -
u_\la b^\a) \,,
\eqno(33)
$$
where $\g \equiv -(u \ w)$ is  analogous to the factor
appearing in the Lorentz transformation for Special
Relativity.

\vskip 2mm

\noindent {\bf B) Evolution equation of the spin $N^\a$ in a three-
dimensional formalism}

\vskip 2mm

We shall now refer $N^\a$, as well as the tangent unit vector
$w^\a$ and the acceleration $a^\a$ of the curve, to the tetrad of
space--time  $\{u^\a, q^\a_i \}$:
$$
\left\{\eqalign{
N^\a &= \hat N^i\,q^\a_i \cr
\vs2
w^\a &= \g\,u^\a + \hat w^i\,q^\a_i \cr
\vs2
a^\a &= - (ua)u^\a + \hat a^i\,q^\a_i \,. \cr
}\right.
\eqno(34)
$$
As mentioned in the introduction, techniques for splitting  space-time have  already been used in previous works by other authors ( Bini, Carini, Massa,...,[6],[13],[18-19], and by Bel, Llosa, Mart\'\i n and Molina [25-26]) following the pioneer work by Cattaneo. From the first expression in (34) we can write  the covariant
derivative of $N^\a$ thus
$$
\frac{\nabla N^\a}{d\tau} = \frac{d \hat N^i}{d\tau} \ q^\a_i +
\hat N^i \ \frac{\nabla q^\a_i}{d\tau} \,,
\eqno(35)
$$
where,  using the second expression
of the decomposition (34), $\dfrac{\nabla q^\a_i}{d\tau}$ is:
$$
\frac{\nabla q^\a_i}{d\tau} = w^\rho\nabla_\rho q^\a_i =
\g\,u^{\rho} \nabla_{\rho} q^\a_i + \hat w^k
q^\rho_k\,\nabla_{\rho} q^\a_i \,.
\eqno(36)
$$
Now using the coefficients (23) of the connection (Ricci's
rotation coefficients) with respect to the above tetrad,  we have:
$$
\eqalign{
 \frac{\nabla q^\a_i}{d\tau} = &-(uw)\[(b q_i)u^\a +
q^\la_i(\S_\la^{\ \a} + \Om_\la^{\ \a})\] \cr
\vs2
&+ \hat w^k\[(\hat\S_{ki} + \hat\Om_{ki})u^\a + \hat\G^j_{ki}
q^\a_j\] \,.
\cr
}
\eqno(37)
$$

With this result and from (32), we can extract from (35) the evolution
equation of the components of $N^\a$ in the connectors frame
$\{q_i^\a\}$:
$$
\ \frac{d\hat N^i}{d\tau}+ \hat\G^i_{kj}\hat w^j \hat N^k = B^i_k
\hat N^k \,,
\quad
\eqno(38a)
$$
with,
$$
\eqalign{
B^i_k \equiv &\ \g(\hat\Om^i_{\ k}-\hat\S^i_{\ k}) \cr
\vs2
&+\frac1{1+\g}\[\hat w^i(\g\hat b_k+\hat a_k) - \hat w_k(\g\hat
b^i+\hat a^i)\] \cr
\vs2
&+\frac1{1+\g}\[\hat w^i\hat w^j(\hat\S_{jk}+\hat\Om_{jk}) -
\hat w_k\hat w^j(\hat\S_j^{\ i}+\hat\Om_j^{\ i})\] \,. \cr
}
\eqno(38b)
$$

Equation (38) describes the evolution of a gyroscope moving
along a curve with 3-velocity $\hat w^i$ with respect to an
arbitrary congruence. As can be seen, the expression involves not
only the geometric objects of the congruence (such as rotation,
deformation, acceleration, or the Zel'manov--Cattaneo connection) but also  the velocity and acceleration of the curve
described by the gyroscope. On the other hand, in general equation (38) is
not a  ``precession" equation, but something more
complicated; this is  because of the presence of $\hat \S_{ij}$ and the
symmetric part of the covariant component of $\hat\G^i_{kj} \hat
w^j$. In this sense, for further applications  it is suitable to write equation (38) in its separate symmetric and antisymmetric parts; i.e.:
$$
\ \frac{d\hat N^i}{d\tau} = ({\cal A}^i_k + {\cal S}^i_k)\hat N^k \,,
\quad
\eqno(39a)
$$
with
$$
\eqalignno{
&\eqalign{
{\cal A}_{jk} \equiv \hat g_{ji}{\cal A}^i_k \equiv &\
\g\,\hat\Om_{jk} +\frac1{1+\g}\[\hat w_j(\g\hat b_k+\hat a_k) -
\hat w_k(\g\hat b_j+\hat a_j)\] \cr
\vs2
&+\frac1{1+\g}\[\hat w_j(\hat\S_{lk}+\hat\Om_{lk}) -
\hat w_k(\hat\S_{lj}+\hat\Om_{lj})\]\hat w^l \cr
\vs2
&+ \frac12(\hat\partial_j\hat g_{kl}-\hat\partial_k\hat g_{jl})w^l
 \,, \cr
} &(39b)\cr
\vspace{3mm}
&{\cal S}_{jk} \equiv \hat g_{ji}{\cal S}^i_k \equiv
-\g\,\hat\S_{jk} - \frac12\hat w^l\hat\partial_l\hat g_{jk} \,. &(39c)
\cr }
$$
Expression (39b) shows a precession rate that generalizes the classical rate of precession appearing in postnewtonian calculations (see, for instance, Weinberg [27]), since the geodetic-precession terms ( Thomas's  gravitational precession) and those refering to the spin-spin interaction (gravitomagnetic effects) are obviously more complicated.

It should be stressed that equation (38) is a tensorial expression on
the quotient manifold since the left hand side is the covariant
derivative of $\hat N^i$ in the sense of the Zel'manov--Cattaneo
connection. However, the evaluation of the angle rotated by the
vector $\hat N^i$ (see Appendix), after one revolution on a closed
orbit, depends on the parametrization of the congruence because this angle 
is defined up to an integer multiple of $2\pi$. As an example, we
mention the simpler case of Minkowski space-time. If we
consider a gyroscope moving on a circular orbit at the
``equatorial" plane and we refer its  evolution to the congruence
defined only by the variation of the time coordinate, which is
irrotational and has no deformation rate, equation (38) is simplified. By solving this equation in Cartesian coordinates ($\hat
\Gamma^i_{jk}= 0$),  in one orbital period the gyroscope precesses by an
angle $2\pi (\g -1)$, which is the correct Thomas precession,
whereas by solving the same equation in spherical coordinates the
angle is $2\pi\g$. It is clear, for  Minkowski space--time, why
the difference betweeen both calculations  of the angle rotated 
is $2\pi$:  the Cartesian connectors evolve by
paralelism, whereas the connectors in spherical coordinates rotate
exactly $2\pi$ after one period of the orbit.

The problem becomes slightly more complicated if we consider, for
example, Schwarzschild or Kerr space--time (also with an
equatorial circular orbit and the congruence defined by $t$
varying alone) because the quotient manifold is a non-flat
3--Riemannian metric, and we therefore have no ``Cartesian
connectors'' that we can use as a reference system.

\vskip 1cm

\noindent {\bf C) Triads of non-natural orthogonal connectors}

\vskip 2mm

In order to avoid ambiguity in the determination of the
rotation of the gyroscope in the general case, it is necessary to
introduce a ``Cartesian-like connectors'' system  on the quotient
space. We first define a general triad of connectors as a
combination of the orthogonal connectors:
$$
h^\a_i = \hat h^k_i q^\a_k \coma {\rm det}(\hat h^k_i) > 0\,,
\eqno(40)
$$
where the coefficients $\hat h^k_i$ only depend on the space
parameters of the congruence -i.e., $\hat h^k_i= \hat h^k_i(z^j)$- 
in order to preserve the meaning given to them. If we now look for
changes in the evolution equation of $N^\a$ (38), we end up with  the
following:
$$
\frac{d\tilde N^i}{d\tau} +
\tilde q^i_j \frac{\hat\nabla\hat h^j_k}{d \tau}=
\tilde B^i_k\tilde N^k\,,
\eqno(41)
$$
where $\hat\nabla$ stands for the covariant derivative with respect
to $\hat\G^i_{jk}$ and $\tilde B^i_k$ denotes the analogous expression to (38b) with respect to the new connectors, i.e,  the ``tilde" is  a
notation for a decomposition similar to (34) but relative to the
new connectors $h^\a_i$. The matrix $\tilde q^i_j$ denotes the
inverse of (40), i. e. $q^\a_k = \tilde q^i_k h^\a_i$.

Obviously, equation (41) can be simplified by assuming that the new
connectors are  transported by paralelism, in the sense of
$\hat\G$, along the curve described by the gyroscope which, as is known, corresponds to spatial Fermi-Walker transport. This assumption implies an
unambiguous criterion for the evaluation of the
gyroscope--orientation change although, as we shall see, for each case it is necessary to evaluate  the holonomy group on the quotient manifold.  At the same time it represents a way to
 intrinsically generalize the connectors associated with Cartesian
coordinates in Minkowski space--time.

With this choice of  the new connectors, and focusing on Born's
congruences ($\tilde\S_{ij}=0$), equation (41) proves to be:
$$
\frac{d\tilde N^i}{d\tau}  = \tilde {\cal A}^i_k\tilde N^k
\quad
\eqno(42a)
$$

$$
\eqalign{
\tilde {\cal A}^i_k \equiv &\ \g\tilde\Om^i_{\ k}
+\frac1{1+\g}\[\tilde w^i\tilde w^j\tilde\Om_{jk}
- \tilde w_k\tilde w^j\tilde\Om_j^{\ i}\] \cr
\vs2
&+\frac1{1+\g}\[\tilde w^i(\g\tilde b_k+\tilde a_k) -
\tilde w_k(\g\tilde b^i+\tilde a^i)\] \,,\cr
}
\eqno(42b)
$$
which is an authentic precession equation for the evolution of the
gyroscope. We shall illustrate this situation in the next Section
with Schwarzschild and the Kerr metrics.

\vskip 1cm

\noindent {\bf 4. GYROSCOPE ORBITING IN SCHWARZSCHILD AND KERR
SPACE--TIME }

\vskip 0.5cm

 \noindent {\bf  A) \ Schwarzschild space-time}

 Let now use Schwarzschild space-time as an example to
calculate the precession of a gyroscope moving in an equatorial
circular orbit. In standard coordinates, this metric is written:
$$
ds^2 = -\(1-\frac{2m}{r}\)dt^2 +
\(1-\frac{2m}{r}\)^{\!\!-1}\!\!dr^2
+ r^2(d\theta^2 + \sin^2\!\!\theta\,d\f^2)\,.
\eqno(43)
$$

On one hand, we have the standard reference time-like congruence:
$$
{\cal C}:\ \left\{\eqalign{
t &= p \cr
\vs2
r &= z^1 \coma \theta = z^2 \coma \f = z^3 \,, \cr
}\right.
\eqno(44)
$$
which has the following quantities associated with it:
$$
\left\{\eqalign{
\xi^\la &= \pd{x^\a}{p} = (1\,,0\,,0\,,0) \cr
\vs2
Q^\la_i &= \pd{x^\a}{z^i} = \delta^\a_i \,,\cr
}\right.
\eqno(45)
$$
\medskip
$$
\xi^2 = 1-\frac{2m}{r} \coma \vf_i =  0 \,,
\eqno(46)
$$
and
$$
\left\{\eqalign{
\hat g_{ij} &={\rm diag}\(\frac1{1-2m/r}\,,\,r^2\,,\,
r^2\sin^2\theta\)  \cr
\vs2
\hat\Om_{ij} &= 0 \coma \hat\S_{ij} = 0 \coma b_i
=\frac{m/r^2}{1-\dfrac{2m}{r}}(1\,,0\,,0) \,. \cr
}\right.
\eqno(47)
$$

On the other hand, we consider a circular orbit in the equatorial
plane with constant angular velocity:
$$
\left\{\eqalign{
t &= p(\tau) \cr
\vs2
r &= R \coma \theta = \frac{\pi}{2}\cr
\vs2
\vf &= \om_\pm\,t  \coma  (\om_\pm = \pm\om\ ,\ \om = {\rm Cte} >
0)\,,
\cr }\right.
\eqno(48)
$$
where $\pm$ denotes the sense of rotation (direct or retrograde) of
the orbit and $p(\tau)$ must be chosen such that $w^\a w_\a = -1$;
i.e.,
$$
 \ w^\a =
\frac{dx^\a}{d\tau}=\frac{1}{X} (1\,,0\,,0\,,\om) \,,
\eqno(49)
$$
where
$$
X \equiv \(\frac{d p}{d \tau}\)^{\!\!-1} =
\(1-\frac{2m}{R}-\om^2\,R^2\)^{\!1/2} \,.
\eqno(50)
$$
Thus, the acceleration of the gyroscope along the orbit is:
$$
a^\a = \frac{\nabla w^\a}{d\tau} =
\frac{\xi^2}{X^2}\(\frac{m}{R^2}-\om^2 R\)\(0\,,1\,,0\,,0\) \,.
\eqno(51)
$$

Since (42b) is a tensorial expression with respect to changes of
connectors triads of the  type shown in (40), and since we are only interested in
evaluating the precession angle of the gyroscope, we can carry out
the calculation by using the connectors associated with the
standard parametrization (44). Thus, we first have:
$$
{\cal A}_{ik} = \frac1{1+\g}\[\hat w_i(\g\hat b_k+\hat a_k) -
\hat w_k(\g\hat b_i+\hat a_i)\] \,,
\eqno(52)
$$
where
$$
\g = -g_{\a\be}u^\a w^\be = \frac{\xi}{X} \,,
\eqno(53)
$$
Therefore, from (47), (49) and (51) we have:
$$
\left\{\eqalign{
{\cal A}_{12} &= {\cal A}_{23} = 0 \cr
\vs2
{\cal A}_{31} &= \mp \frac{\om R}{\xi\,X^2}\[1-\frac{3m}{R} -
\xi\,X\] \,.
\cr
}\right.
\eqno(54)
$$

The precession angle and the sense of rotation can be obtained from the
following dual vector (see Appendix):
$$
\Om^i = -\frac12\frac1{\sqrt{\hat g}}\epsilon^{ijk} {\cal A}_{jk} =
-\frac1{\sqrt{\hat g}} {\cal A}_{31}\,\delta^i_2 \,,
\eqno(55)
$$
which turns out to be:
$$
 \ \Om^i = \pm \frac{\om}{R\,X^2}\[1-\frac{3m}{R} -
\xi\,X\]\,(0\,,1\,,0) \,.
\eqno(56)
$$

$\bullet$ By taking in (56) the orbit to be geodesic ($\om^2 =
m/R^3$), we have:

$$
\Om^\theta = \mp \frac{\sqrt{m/R}}{R^2\,X_g}\[\sqrt{1-\frac{2m}{R}} -
\sqrt{1-\frac{3m}{R}}\,\] \cases{< 0 \cr > 0 \,,\cr}
\eqno(57)
$$
where $X_g$ stands for the value of $X$ when the orbit is geodesic
($X_g \equiv \sqrt{1-3m/R}$). Hence, the precession is ${\it
direct}$ or ${\it retrograde}$ for ${\it direct}$  or ${\it
retrograde}$ orbits respectively,  and the rotated angle  after one 
period of proper time is:
$$
\Delta\a = \pm\Om\,\frac{2\pi\,X_g}{\om} =\pm
2\pi\[\sqrt{1-\frac{2m}{R}} - \sqrt{1-\frac{3m}{R}}\,\] \,,
\eqno(58)
$$
where the corresponding sign is considered for the respective sense
of the orbit.

$\bullet$ On other hand, by taking $m=0$ in  expression (56) we
can recover the result for Minkowski space-time. As is known,
for the  case of a ${\it direct}$ orbit the precession is ${\it
retrograde}$ and vice-versa,  which is clear from the
component:
$$
\Om^\theta = \frac{\pm \om}{R\,(1-\om^2\,R^2)}\[1-
\sqrt{1-\om^2\,R^2}\,\] \cases{> 0 \cr < 0 \,.\cr}
\eqno(59)
$$
The rotated angle is:
$$
\Delta\a \equiv \mp \Om\,\frac{2\pi\,X_m}{\om}
=\mp 2\pi\[(1-\om^2\,R^2)^{-1/2} - 1\]\,,
\eqno(60)
$$
$X_m$ being  the value of $X$ (50) for $m=0$, and the signs
$\mp$ standing for ${\it direct}$ or ${\it retrograde}$ orbits,
respectively.

As can be checked, results (58) and (60) do not correspond to
the expressions obtained by Rindler and Perlick [2]. The
expressions obtained by these authors  come simply from adding an
angle of $2 \pi$ to the result obtained if the congruence defined by all the circular orbits centered in the symmetry axes and  ``orthogonal" to it is considered (in this way the matrix $B^i_k$ of (38) is  reduced to $\hat\Om^i_{\ k}$ and 
$\hat\G^i_{kj} = 0$). This procedure is quite reasonable for simple examples, but it does not seem to be supported in the general case and, moreover, it only gives the rotated angle  but allows no conclusions to be drawn about the influence of fields  associated with  some suitable reference congruence.  In any case, the question is as  follows: what 
criterion is used to say that the  precession angle is any given angle?.
A possible answer for this question comes from the use of  a triad
of connectors transported by paralelism in the sense of $\hat
g_{ij}$, as we have shown, although obviously there is no unique
answer and some other similarly reasonable criterion may be used.

With respect to the difference between  results (58), (60) and
those obtained by Rindler and Perlick, it should be noted that this is due to the fact that the connectors transported by
parallelism along a closed curve in a Riemannian space rotate at a 
certain angle (holonomy), which is zero in a flat Minkwosky
space-time. Therefore,  we rely on a procedure suitable for calculating the
``total" angle that the gyroscope rotates by adding the 
angle rotated by the parallel transported connectors to expressions
(58) and (60). To evaluate the angle rotated, after one
revolution, by the triad transported by parallelism, we avoid 
possible ambiguity by forcing the angle to be zero at the
Minkowskian limit.

To this end, let us consider  the parallel transport  for
the connectors in the quotient space of Schwarzschild space-time.
The final equation is:
$$
\frac{d q^i}{d \tau} = P^i_{\ j} q^j\,,
\eqno(61)
$$
with
$$
P^i_{\ j} \, : \quad P^i \equiv - \frac 12 \frac{1}{\sqrt{\hat g}}
\epsilon^{ijk}\hat g_{jl} P^l_{\ k}= \pm
\frac{\om}{R} \frac{\sqrt{1-\frac{2m}{R}}}
{\sqrt{1-\frac{3m}{R}}}\,(0,1,0)\,.
\eqno(62)
$$
By using the techniques shown in the Appendix, it is trivial to conclude that the connector rotates, after a period $T_p$, by an angle:
$$
|\Delta \beta | \equiv  \frac{2 \pi}{T_p}\,(P^i P_i)^{1/2}=
\Bigg |2 \pi\sqrt{1-\frac{2 m}{R}} + 2k\pi\Bigg| \,.
\eqno(63)
$$
Since this angle must be zero at the Minkowskian limit, we
have that the corresponding angles for the respective ${\it
direct}$ and ${\it retrograde}$ orbit cases are:
$$
\Delta \beta = \pm 2 \pi \[1 -\sqrt{1-\frac{2m}{R}}\]\,.
\eqno(64)
$$
Henceforth, the precession for the geodesic case can be corrected
as follows:
$$
\Delta\a +  \Delta\be = \pm  2\pi \(1-\sqrt{1-\frac{3m}{R}}\)\,,
\eqno(65)
$$
which is consistent with the result obtained by Rindler and Perlick.

\vskip 5mm

\noindent {\bf B) \ Kerr space-time}

Finally, we should like to complete this analysis by showing the
results obtained  for  Kerr space-time.

We first calculate the precession angle of the gyroscope by using
the connectors associated with the standard parametrization (44) of
the reference time-like congruence, as well as the  circular orbit on the
equatorial plane (48), $\{t, r, \theta, \vf \}$ being 
Boyer-Lindquist coordinates. In this case, if the orbit is
a geodesic, the constant angular velocity turns out to be:
$$
\om_{\pm} = \frac{\om_s}{a \om_s \pm 1}\,,
\eqno(66)
$$
with $\om_s \equiv
+\sqrt{m/R^3}$ and where the respective sign shows that  the orbit is  ${\it direct}$ or ${\it retrograde}$.

It is straightforward to calculate that  (55) becomes:
$$
\Om^\theta = \mp \frac{\om_s^2}{R \xi^3 X_k} C \,,
\eqno(67)
$$
where

$$
\left\{\eqalign{
C &\equiv \displaystyle{ -a(a \om_s \pm \xi^2) \pm
\displaystyle{\frac{ \om_s \xi^2 R^2 (\xi^2+a^2/R^2)}{a \om_s \pm
\xi^2 \pm \xi X_{\kappa}}}} \, > 0 \cr
\vs2
\xi &\equiv \sqrt{1-2m/R} \cr
\vs2
X_k &\equiv  \sqrt{1-3 m/R \pm 2a\om_s} \,.\cr
}\right.
\eqno(68)
$$
The angle rotated after one revolution (with its corresponding sign
for the respective {\it direct} or {\it retrograde} orbit)  turns
out to be:

$$
\Delta \a = \pm 2 \pi \frac{\om_s}{\xi^3}
\[ -a(a \om_s \pm \xi^2) \pm \frac{R^2 \om_s \xi^2
(\xi^2+a^2/R^2)}{a \om_s \pm \xi^2 \pm \xi X_k} \]\,.
\eqno(69)
$$
It can be readily checked that the reduction $a=0$ provides the
results obtained for the Schwarzschild case. As we did for the 
Schwarzschild case, the evaluation of the parallel transport for
the connectors in the quotient metric of Kerr space-time leads to
an equation like (61), with:
$$
P^i = \pm \frac{\om_s (\xi^4-\om_s^2 a^2)}{X_k \xi^3 R}\,
(0,1,0)\,.
\eqno(70)
$$
And the angle rotated by the connectors (with the good Minkowskian limit) is:
$$
\Delta \be = \pm 2 \pi \[1 -\frac{(\xi^4-\om_s^2 a^2)}{\xi^3} \]\,.
\eqno(71)
$$
Since we have that $ \om_s \, C -(\xi^4-\om_s^2 a^2) =
-X_k \xi^3$, we thus have  that the total precession angle is:
$$
\Delta \a + \Delta \be = \mp 2 \pi \[  \sqrt{1-3 m/R \pm 2 a
\om_s} -1 \]\,.
\eqno(72)
$$

\vskip 1cm

\noindent {\bf APPENDIX}

\vskip 0.5cm

In this Appendix we wish to show the fundamental aspects of a differential precession equation on a three-dimensional Riemannian manifold (${\cal V}_3,g_{ij}$); that is, a differential equation of the following type:
$$
\dt{N^i}{\tau} = {\cal A}^i_k(\tau)\,N^k \coma
{\cal A}_{jk}\equiv g_{ij}{\cal A}^i_k = -{\cal A}_{kj}\,.
\eqno(A1)
$$
The general solution of this kind of equation is:
$$
 N^i(\tau) = \F^i_k(\tau)\,N^k_{_{(0)}}\,,
\eqno(A2)
$$
where
$$
\left\{\eqalign{
&\dot \F^i_j = {\cal A}^i_k(\tau)\,\F^k_j \cr
\vs2
& \F^i_j(0) = \de^i_j \,. \cr
}\right.
\eqno(A3)
$$

Although it is already quite well known, we show  that (A1) is
an authentic equation of precession. Indeed, by introducing the
dual vector (up to a sign) of ${\cal A}_{jk}$ in the sense of
$g_{ij}$
$$
\Om^i \equiv - \frac12\,\eta^{ijk}\,{\cal A}_{jk} \coma
\eta^{ijk}=\frac1{\sqrt{g}}\epsilon^{ijk}\,,
\eqno(A4)
$$
we see that:
$$
{\cal A}^i_kN^k = \eta^{ijk}\,\Om_j N_k \equiv
+\,(\vec\Om\ext\vec N \,)^i\,,
\eqno(A5)
$$
where the vectorial product is also understood in the sense of
$g_{ij}$.

Furthermore, the matrix ${\cal A}^i_k$ has the following and
well-known interesting property (because of the antisymmetric
character of ${\cal A}_{jk}$):
$$
{\cal A}^i_j\,{\cal A}^j_k\,{\cal A}^k_l = -\Om^2\,{\cal A}^i_l
\coma ({\cal A}^3 = -\Om^2\,{\cal A})\,,
\quad
\eqno(A6)
$$
where
$$
\Om^2 \equiv \Om^i\,\Om_i = -\frac12 {\cal A}^i_j\, {\cal A}^j_i
\equiv -\frac12 {\rm tr}\,{\cal A}^2 \,.
\eqno(A7)
$$

Let us now assume that all components of ${\cal A}_{ij}$ are {\it
constants}. Then, the general solution for (A3) is:
$$
\F(\tau) = e^{{\cal A}\,\tau}\,.
\eqno(A8)
$$
By virtue of (A6), we have:
$$
 \F(\tau) = I + \frac{\sin\Om \tau}{\Om}\,{\cal A} +
\frac{1-\cos\Om \tau}{\Om^2}\,{\cal A}^2\,,
\eqno(A9)
$$
and the solution of (A1) can therefore  be written as:
$$
\eqalign{
\vec N(\tau) = \vec N_0 &+ \frac{\sin\Om
\tau}{\Om}\,\,
\vec\Om\ext\vec N_0 \cr
\vs2
&+ \frac{1-\cos\Om \tau}
{\Om^2}\,\,\vec\Om\ext(\vec\Om\ext\vec N_0) \,, \cr
}
\eqno(A10)
$$
which can be simplified to give:
$$
\eqalign{
\vec N(\tau) = (\vec n \cdot\vec N_0)\,\vec n &+
\sin\Om\tau\,(\vec n\ext\vec N_0)\cr
\vs2
 &- \cos\Om \tau\,\[(\vec n \cdot\vec N_0)
\,\vec n -\,\vec N_0\] \,,
\cr }\quad
\eqno(A11)
$$
with
$$
\vec n \equiv \frac{{\vec\Om}}{\Om} \coma \vec n \cdot
\vec N_0\equiv g_{ik} n^i N_0^k \,.
\eqno(A12)
$$

This expression shows that the angle between $\vec \Om$ and
$\vec N$ remains unchanged:
$$
\vec n\cdot\vec N(\tau) = \vec n\cdot\vec N_0
\coma
\forall \tau
\eqno(A13)
$$
So, if we take for instance  $\vec n\cdot\vec N_0 = 0$,
this leads to the final expression:
$$
\vec N(\tau) = N_0\,\(\cos\Om \tau \  \vec n_1 +
\sin\Om \tau \ \vec n_2\) \,,
\eqno(A14)
$$
with $\{\vec n_1, \vec n_2, \vec n_3\}$ a triad of orthonormal
vectors with the following notation:
$$
\left\{\eqalign{
&\vec n_1 \equiv \frac{\vec N_0}{N_0} \cr
\vs2
&\vec n_2 \equiv \vec n_3\ext\vec n_1 \cr
\vs2
&\vec n_3\equiv\vec n \,.
\cr }\right.
\eqno(A15)
$$

Finally, it is clear from above, (A14), that the {\it precession
angle} after a value of the parameter $\tau$ is equal to:
$$
\Delta\a = \Om\,\tau \,,
\eqno(A16)
$$
which is a positive defined quantity because $\Om$ is the norm of
the dual vector $\Om^i$. Nevertheless, in practice it is useful to assign a sign (positive or negative) to this angle, as done in this article,  according to whether the rotation defined by the vector  $\vec \Om$ (from $\vec n_1$ to $\vec n_2$) is {\it direct} or {\it retrograde} for each particular context.

\vskip 1cm

\noindent {\bf REFERENCES}

\vskip 2mm

\noindent [1] {\it Gravity Probe B Mission} web page,
http://einstein.stanford.edu/

\noindent [2] Rindler, W. and Perlick, V., (1990). \ {\it Gen. Rel.
Grav.}, {\bf 22}, 1067.

\noindent [3] Rindler, W. (1977). \ {\it Essential Relativity},
(Springer-Verlag, Berlin) 2nd. ed.

\noindent [4] Iyer, B. R. and  Vishveshwara, C. V., (1993). \ {\it
Physical Rev. D}, {\bf 48}, 5706.

\noindent [5] Shapiro, I. I., Reasenberg, R. D., Chandler, J. F.,
and Babcock, R. W., (1988).  \ {\it Physical Rev. Lett.}, {\bf
61}, 2643.

\noindent [6] Bini, D., Gemelli, G. and Ruffini, R. , (2000).  \
{\it Physical Rev. D}, {\bf 61}, 064013.

\noindent [7] Herrera, L., Pavia, F. and Santos, N.O. , (2000). \
{\it Class. Quantum Grav.}, {\bf 17}, 1549.

\noindent [8] Herrera, L. and Hern\'andez-Pastora, J.L. , (2000).
 \ {\it Journal of Mathematical Physics},  {\bf 41}, 7544.

\noindent [9] Thomas, L. H.,  (1926). \ {\it Nature}, {\bf 117},
514.

\noindent [10] Fokker, A. D.,  (1920). \ {\it Kon. Akad. Weten.
Amsterdam, Proc.}, {\bf 23}, 729.

\noindent [11] de Sitter, W., (1920). \ {\it Mon. Not. Roy.
Astron. Soc.}, {\bf 77}, 155.

\noindent [12] Shiff, L. I., (1960).  \ {\it Physical Rev. Lett.},
{\bf 4}, 215.

\noindent [13] Massa, E. and Zordan, C. (1975), \ {\it Relative Kinematics in General Relativity: The Thomas and Fokker precessions. Meccanica}
{\bf 10}, 27.

\noindent [14]  Cattaneo, C., (1958), \ {\it Il Nuovo Cimento}
{\bf 10}, 318.

\noindent [15]  Cattaneo, C., (1959), \ {\it Il Nuovo Cimento}
{\bf 11}, 733.

\noindent [16]  Cattaneo, C., (1959), \ {\it Ann. dii Mat. Pura ed
Appl}, S. IV. T. XLVIII, 361.

\noindent [17]  Cattaneo, C., (1963), \ {\it Comptes Rendus de l'Acad. des Sc.}
{\bf 256}, 3974.

\noindent [18]  Jantzen, R. T., Carini, P. and Bini, D. (1992), \ {\it Annals of Physics}
{\bf 215}, 1 (section IX).

\noindent [19]   Bini, D. and de Felice, F., (2000), \ {\it Classical Quantum Grav.} {\bf 17}, 4627.

\noindent [20]  Bel, Ll., Mart\'\i n, J. (1980), \ {\it Ann. Inst. Henri Poincar\'e} {\bf 33-A}, 409.

\noindent [21]  Zel'manov A., (1956), \ {\it Sov. Phys. Dokl.}
{\bf 1}, 227.

\noindent [22]  Cattaneo-Gasparini,I., (1961), \ {\it C. R. Acad.
Sc.} {\bf 252}, 3722.

\noindent [23] Papapetrou, A., (1951), \ {\it Proc. Royal Soc. (London)} {\bf 209 A}, 248

\noindent [24] Soffel, M. H., (1989), \ {\it Relativity in
Astrometry Celestial Mechanics and Geodesy}. Springer-Verlag.
Berlin.

\noindent [25]  Bel, Ll., and Llosa, J., (1995), \ {\it General Relativity and Gravitation} {\bf 27}, 1089

\noindent [26] Bel, Ll., Mart\'\i n, J. and Molina, A., (1994), \ {\it Journal of the Physical Society of Japan} {\bf 63}, 4350

\noindent [27] Weinberg, S.,  (1972), \ {\it Gravitation and Cosmology: Principles and Applications of the General Theory of Relativity}. Wiley, New York.
\bye